\documentclass[twocolumn,
aps,nofootinbib,showpacs,showkeys,preprint
tightenlines
] {revtex4-1}

\usepackage{epsf,epsfig,subfigure,graphicx,amsmath,amssymb}
\usepackage{color}
\usepackage{float}
\newcommand{\dis}[1]{\begin{equation}\begin{split}#1\end{split}\end{equation}}

\newcommand{\gev}{\,\textrm{GeV}}

\newcommand{\ie}{{\it i.e.}\ }

\newcommand{\Vckm}{$V_{\rm CKM}$}

\newcommand{\Z}{{\bf Z}}

\begin{document}

\title{\large\bf  The CKM matrix with maximal CP violation from $\Z_{12}$ symmetry}

\author{Jihn E. Kim
\email{jekim@ctp.snu.ac.kr}
}
\affiliation{
GIST College, Gwangju Institute of Science and Technology,  Gwangju 500-712, Korea \\
 }
%\affiliation{
%Department of Physics and Astronomy and Center for Theoretical Physics, Seoul National University, Seoul 151-747, Korea
%}
\begin{abstract}
The recent accurate determination of the CKM parameters including the maximal CP phase  $\frac{\pi}{2}$ enables us to write down the up-type and down-type quark mass matrices to a  high degree of accuracy. The lightest element($\bar uu$ element) of the quark mass texture (not the mass eigenvalue) has a power $\lambda^6$ where $\lambda=\sin\theta_C$. The CP phase of $2\pi$ divided by an integer hints a discrete symmetry. Since $\lambda^6$ is the highest power of $\lambda$ among the quark mass matrix elements, we present as an example a possibility that the maximal CP phase  $\frac{\pi}{2}$ is obtainable from a supersymmetric $\Z_{12}$ discrete symmetry model.
\end{abstract}


\maketitle

%%%%%%%%%%%%%%%%%%%%%%%%%%%%%%%%%%%%%%%%%%%%%%%%%%%%%%%%%%%%
\section{Introduction}\label{sec:Introduction}
%%%%%%%%%%%%%%%%%%%%%%%%%%%%%%%%%%%%%%%%%%%%%%%%%%%%%%

The Cabibbo-Kobayashi-Maskawa(CKM) matrix has proved to be an effective description of the weak CP violation where the observed weak CP phase seems to be maximal \cite{CKM73}. The maximality of the weak CP violation has been pointed out by a number of authors from the known quark mass ratios \cite{Shin84}. If it is maximal, there is a hope to obtain it from the dynamical details of the mass generation mechanism \cite{MassGenFirst}.

The maximal CP violation is an attractive idea from the theoretical point of view. For example, from U(1)$^3$ symmetry, Georgi, Nelson and Shin tried to obtain the phase $\frac{\pi}2$ which however has led to an unacceptably large generation of the vacuum angle $\theta_{\rm weak}$ \cite{GeorgiShin85}.
The CKM matrix can be expanded in terms of a parameter $\lambda$ which has been suggested as a superheavy mass ratio by Froggatt and Nielsen(FN) \cite{FN79}. This superheavy mass ratio is typically provided by a superheavy vacuum expectation value(VEV) of the standard model(SM) singlet field compared to the Planck scale $M_P\simeq 2.44\times 10^{18}\gev$. In view of the recent accurate determination of the CKM matrix elements in the last decade  \cite{PData10}, now time is ripe enough to scrutinize these old ideas whether they are realizable or not.

The well-known facts on the CKM matrix are
\begin{itemize}
\item[A.] To read the weak CP violation directly from the \Vckm~ elements,  Det.\Vckm~ is better to be real \cite{KimSeo11}. Note however that the phase of Det.\Vckm~ is not observable. To have the weak CP violation, \Vckm~  itself must be complex.

\item[B.] If any among the nine elements of \Vckm~ is zero, then there is no
      weak CP violation.% \cite{FritzschOneZero}.
\item[C.] The Cabibbo angle $\sin\theta_C=\lambda$ is a good expansion parameter \cite{Wol83}.

\item[D.] With Item A satisfied, the product $V_{31}V_{22}V_{13}$ is the barometer of weak CP violation \cite{KimSeo11}.

\item[E.] \Vckm~ is derivable from the Yukawa texture \cite{MassGenFirst}.

\end{itemize}

In Ref. \cite{KimSeo11}, an exact CKM matrix satisfying the above and containing the approximate Wolfenstein form was presented. In this exact form with Det.\Vckm=1, vanishing of any one parameter among $\theta_{1,2,3}$ or $\delta$ makes \Vckm~ real.  The CP phase $\delta$ has been determined to be almost maximal, $\delta= 89.0^{\rm o}\pm 4.4^{\rm o}$ \cite{PData10}. On the other hand, in Wolfenstein's approximate form there are two phases, the phase $\delta_b$ of the (13) element and the phase $\delta_t$ of the (31) element. These phases are proportional to the imaginary parameter $i\eta$ of Wolfenstein \cite{Wol83} but the physically observable CP phase is $\delta_b+\delta_t$ which is our single parameter $\delta$.\footnote{This was observed also in \cite{Ahn11}. See also \cite{QinMaPLB11}.}
As shown in \cite{KimSeo11}, Item D shows the weak CP violation directly from each element of
\Vckm. The weak CP invariant quantity the Jarlskog determinant \cite{Jarlskog85} removes the real part in the expression of Ref. \cite{KimSeo11} and leaves only the imaginary part. Therefore, Item A is simpler than calculating the Jarlskog determinant. Among 6 contributions to Det.\Vckm, any one is a good barometer of CP violation. In particular, the product of the skew diagonal elements (Item D) is a quick barometer of the weak CP violation.

The largest error, numerically 0.0011, in the currently determined matrix elements resides in \Vckm$_{(23)}$ and \Vckm$_{(32)}$. It is of order $\lambda^3$ which is an order $\lambda$ smaller than the leading terms of  \Vckm$_{(23)}$ and \Vckm$_{(32)}$, and hence the leading terms in each element of \Vckm~ are pretty well determined by now.  Since the $\lambda$ expansion of \Vckm~ has determined all the elements accurately, with the current knowledge of the quark mass ratios \cite{PData10}  it is fairly straightforward to calculate the quark mass matrices in the  weak basis, $\tilde M^{u}$ and  $\tilde M^{d}$ \cite{KimSeo11}. Now, we can attack the problem posed in Item E \cite{MassGenFirst}.

Toward the Yukawa textures, the most obvious try would be U(1) symmetries \cite{GeorgiShin85}. If they are gauged, the anomaly cancelation should be satisfied. Here, however, we attempt to introduce a discrete symmetry toward the Yukawa textures \cite{KimSeo10}, in particular a $\Z_{12}$ symmetry. The CKM fitting with up and down quark mass matrices allows the smallest entry, \ie the (11) entry of the up quark mass matrix, of ${\cal O}(\lambda^6)$ and hence $\Z_{12}$ is tried in this paper. For discrete symmetries, it is better for them to be of discrete gauge symmetry \cite{Krauss89}. Here, however, we do not satisfy the discrete gauge symmetry at the Planck scale. We anticipate that the gravitational interaction breaks the discrete symmetry, but the gravitational interaction, breaking the discrete symmetry, respects the flavor independence \cite{FritzschDemo90} since the quark masses are much smaller than the Planck scale $M_P$. Far below $M_P$, the nongravitational interaction respects the discrete symmetry we propose here.

In Sec. \ref{sec:Qmasses}, we parametrize the quark masses as powers of $\lambda$. From the known CKM matrix, we identify the left-hand and right-hand unitary matrices for diagonalization of quark mass matrices. If the quark mass matrices are given, these unitary matrices are determined. So, our choices are confined to Hermitian mass matrices. Specifying the left-- and right--unitary matrices is a kind of a solution of an inverse problem. In Sec. \ref{sec:YukTexture}, we obtain the Yukawa texture and introduce $\Z_{12},\, \Z_{4}$ and $\Z_{3}$ discrete symmetries. In Sec. \ref{sec:MaxCP}, we introduce supersymmetry(SUSY) and attempt to obtain the CP phase $\frac{\pi}2$ from the allowed superpotential terms. Sec. \ref{sec:Conclusion} is a conclusion.

%%%%%%%%%%%%%%%%%%%%%%%%%%%%%%%%%%%%%%%%%%%%%%%%%%%%%%%%%%%%
\section{Quark mass matrices}\label{sec:Qmasses}

To determine the quark mass matrices as accurately as possible, it is necessary to have
the Wolfenstein parametrizations valid up to high orders of $\lambda$. In Ref.  \cite{KimSeo11}, the $\lambda$ expansion was obtained up to ${\cal O}(\lambda^6)$,
\begin{widetext}
\dis{
\left(\begin{array}{lll}  1-\frac{\lambda^2}{2}-\frac{\lambda^4}{8}-\frac{\lambda^6}{16}(1+8\kappa_b^2), \quad &\lambda  ,   &  \lambda^3 \kappa_b\left(1+\frac{\lambda^2}{3}\right) \\ [1em]
-\lambda+\frac{\lambda^5}{2}(\kappa_t^2-\kappa_b^2) ,\quad
 & \begin{array}{l}
1-\frac{\lambda^2}{2}-\frac{\lambda^4}{8}-\frac{\lambda^6}{16} \\[0.2em]
-\frac{\lambda^4}{2}(\kappa_t^2+\kappa_b^2-2\kappa_b\kappa_t e^{-i\delta})\\[0.2em]
-\frac{\lambda^6}{12}\left(7 \kappa_b^2+\kappa_t^2-8\kappa_t \kappa_b e^{-i \delta}\right)
\end{array},      &
\begin{array}{l}
\lambda^2\left(\kappa_b-\kappa_t e^{-i\delta} \right) \\[0.2em]
 -\frac{\lambda^4}{6}(2\kappa_t e^{-i \delta}+\kappa_b)
\end{array}\\ [2.5em]
      -\lambda^3 \kappa_t e^{i\delta}\left(1+\frac{\lambda^2}{3}\right) ,  &
\begin{array}{l}
 -\lambda^2\left(\kappa_b-\kappa_t e^{i\delta} \right)\\[0.2em]
 -\frac{\lambda^4}{6}(2\kappa_b+ \kappa_te^{i\delta})
\end{array}
 , &
\begin{array}{c}
1-\frac{\lambda^4}{2}(\kappa_t^2+\kappa_b^2
      -2\kappa_b\kappa_t e^{i\delta})\\[0.2em]
-\frac{\lambda^6}{6} \left(2[\kappa_b^2+\kappa_t^2]-\kappa_t \kappa_b e^{i \delta}
\right)
\end{array}
\end{array}\right) \label{eq:KSrotated}
}
\end{widetext}
where
$\lambda=0.22527\pm 0.00092, \kappa_t=0.7349\pm 0.0141,~~\kappa_b=0.3833\pm 0.0388$, and
$\delta=89.0^{\rm o}\pm 4.4^{\rm o}$.

Under the mass eigenstate bases $u^{\rm (mass)}=(u,c,t)^T$ and $d^{\rm (mass)}=(d,s,b)^T$, the observed quark masses are
\dis{
 \frac{M^{(u)}}{m_t}=\left(\begin{array}{ccc}  \lambda^7 u   &0 ,   &  0 \\
      0 &    \lambda^4 c& 0 \\
       0 &  0&1
\end{array}\right), \
\frac{M^{(d)}}{m_b}=\left(\begin{array}{ccc}  \lambda^4 d   &0 ,   &  0 \\
      0 &    \lambda^2 s& 0 \\
       0 &  0&1
\end{array}\right) \label{eq:MuMd} }
where $u,c,d$ and $s$ are four real parameters of ${\cal O}(1)$ \cite{PData10},
\dis{
&u= 0.50^{+0.16}_{-0.13},~ c= 2.8\pm 0.2, \\
&d= 0.45^{+0.10}_{-0.08},~ s=0.49\pm 0.13.
 }
These are ${\cal O}(1)$ parameters but $c$ is about $1/\lambda$ times larger than the others. Even though this is a peculiarity, we use this form so that the second family $c$ element of Eq. (\ref{eq:MuMd}) is an even power of $\lambda$. For the first family member $u$, using $\lambda^7$ or $\lambda^8$ does not matter since the parameter $u$ does not appear as an important term in the determinant. If we used $c$ as the coefficient of $\lambda^3$, we do not achieve the nice features of the present model discussed below. So, we speculate that $u,c,d,$ and $s$ are determined by another mechanism, probably by topological numbers of the internal space rather than the VEVs of the FN fields.
The mass matrices in the weak eigenstate bases are related to the above by bi-unitary transformations,
\dis{
\tilde M^{(u)}=  R^{(u)\dagger}  M^{(u)}L^{(u)}  ,\quad \tilde M^{(d)}=  R^{(d)\dagger} M^{(d)} L^{(d)}
}
where $R^{(u),(d)}$ and $L^{(u),(d)}$ are unitary matrices used for the R-handed and L-handed quark fields.

Actually, obtaining the specific forms of mass matrices from \Vckm\ is a kind of an inverse problem, needing the information on the unitary matrices diagonalizing mass matrices. So, there are two ambiguities: firstly the right handed unitary matrices are arbitrary and second even in this case the left handed unitary matrices have many possibilities. We will choose the unitary matrices so that many zeros appear in the left-handed matrices. The CKM matrix can be represented as
    \dis{& V_{\rm CKM}= L^{(u)} L^{(d)\dagger}=\left(\begin{array}{ccc}  1&0&0\\[1 em]
0& 1 & 0\\ [1em] 0& 0 &e^{i \delta} \end{array}\right)
\left(
\begin{array}{ccc}  1&0&0\\[1 em]0& c_2 &s_2\\ [1em]
0& -s_2 &c_2 \end{array}\right)\\
&~\times
\left( \begin{array}{ccc} c_1&s_1&0\\[1 em]
-s_1& c_1 & 0\\ [1em] 0& 0 &1  \end{array}\right)
\left( \begin{array}{ccc}  1&0&0\\[1 em] 0& c_3 & -e^{i \delta}s_3\\ [1em]
0& e^{i \delta}s_3 &c_3  \end{array}\right)
\left( \begin{array}{ccc}  1&0&0\\[1 em]
0& 1 & 0\\ [1em]0& 0 &e^{-i \delta}  \end{array}\right).\label{eq:CKMfivefact}
  }
Of course, Eq. (\ref{eq:CKMfivefact}) is one among many published forms in the literature \cite{eq:CKMfactors}.
Since  Eq. (\ref{eq:CKMfivefact}) is composed of a product of five matrices, $L^{(u)}$ and $L^{(d)}$ can take different forms. Now, let us choose the left hand matrices such that many zeros appear in $L^{(d)}$,
\dis{L^{(u)}=\left(\begin{array}{ccc}  1&0&0\\[1 em]
0& 1 & 0\\ [1em]0& 0 &e^{i \delta}  \end{array}\right)
\left(\begin{array}{ccc}  1&0&0\\[1 em]0& c_2 &s_2\\ [1em]
0& -s_2 &c_2  \end{array}\right)
\left(\begin{array}{ccc} c_1&s_1&0\\[1 em]
-s_1& c_1 & 0\\ [1em]0& 0 &1  \end{array}\right)}
and
\dis{L^{(d) \dagger}=\left(\begin{array}{ccc}  1&0&0\\[1 em]
0& c_3 & -e^{i \delta}s_3\\ [1em]0& e^{i \delta}s_3 &c_3  \end{array}\right)
\left(\begin{array}{ccc}  1&0&0\\[1 em]0& 1 & 0\\ [1em]
0& 0 &e^{-i \delta}  \end{array}\right).
}
Then, for $R^{(u),(d)}=L^{(u),(d)}$,\footnote{The mass matrices $\tilde M^{(u)}$ and $\tilde M^{(d)}$ must be hermitian.}   $\tilde M^{(d)}$  contain four zeros
\begin{widetext}
\dis{
\tilde M^{(u)}&=\left(\begin{array}{ccc}  (c+\kappa_t^2 \lambda) \lambda^6 ,  & -(c+\kappa_t^2 ) \lambda^5 ,
 &  \kappa_t\lambda^3 (1+\frac{1}{3}\lambda^2) \\[0.2em]
  -(c+\kappa_t^2 ) \lambda^5 ,&  c\lambda^4(1- \frac{1}{3} \lambda^2), & -\kappa_t \lambda^2+\frac{\kappa_t}{6}\lambda^4 +O(\lambda^6) \\[0.2em]
       \kappa_t\lambda^3 (1+\frac{1}{3}\lambda^2), &-\kappa_t \lambda^2+\frac{\kappa_t}{6}\lambda^4 +O(\lambda^6) ,
       & 1-\kappa_t^2\frac{\lambda^4}{2}-\kappa_t^2\frac{\lambda^6}{3}
\end{array}\right)\\[0.4em]
  \tilde M^{(d)}&=\left(\begin{array}{ccc}  d \lambda^4(1+\frac{2}{3}\lambda^2),   & 0  , &  0 \\[0.2em]
      0 ,&  s\lambda^2+(\kappa_b+\frac{s}{3})\lambda^4+(\frac{8}{45}s+\frac{2\kappa_b^2}{3})\lambda^6,
       &  \kappa_b e^{i \delta} (-\lambda^2+(s-\frac{1}{3})\lambda^4) +O(\lambda^6)  \\[0.2em]
       0, &  \kappa_b e^{-i \delta} (-\lambda^2+[s-\frac{1}{3}]\lambda^4) +O(\lambda^6),
       & 1-\kappa_b^2\lambda^4+\kappa_b^2(s-\frac{2}{3})\lambda^6
\end{array}\right).             \label{eq:TextureRL}
}
\end{widetext}
which will be used below. Note that there appear four zeros in $\tilde M^{(d)}$.

%%%%%%%%%%%%%%%%%%%%%%%%%%%%%%%%%%%%%%%%%%%%%%%%%%%%%%%%%%%%%
\section{Yukawa textures}\label{sec:YukTexture}
The maximality of CP violation can be related to the Yukawa texture. The $\lambda$ expansion may come from the FN  mechanism \cite{FN79}. So far, the FN mechanism is mostly applied to continuous symmetries. Here, we attempt to obtain the texture (\ref{eq:TextureRL}) using discrete symmetries \cite{KimSeo10}, which may be useful in determining the CP phase. Since the observed CP phase seems maximal $\delta\simeq\frac\pi{2}$, discrete symmetries might have worked in determining it. Because the highest power of $\lambda$ in Eq. (\ref{eq:TextureRL}) is ${\cal O}(\lambda^6)$, here we choose the discrete symmetry $\Z_{12}$: $n\pm 12$ is identified with $n$. Since we will try $|X_{\pm 1}^{6+a}|=|X_{\pm 1}^{6-a}|$, $\Z_{12}$ is the discrete symmetry we need.

To facilitate the algebra and also toward the gauge hierarchy solution, we introduce $N=1$ supersymmetry (SUSY). With SUSY, we can follow the set-and-forget principle in the superpotential $W$. But that is a fine-tuning in a sense, and hence we try to introduce more symmetries to obtain Eq. (\ref{eq:TextureRL}) naturally.

Let us introduce two Higgs doublets $H_u$ and $H_d$, and several GUT scale singlets $X_1^{d,u}, X_{-1}^{d,u}, X_{6}^{d,u}$ and $X_{0}^{d,u}$. The fields with indices $u$ give mass to up-type quarks and the fields with indices $d$  give mass to down-type quarks. The $H_u$ and $H_d$ fields coupling to the up-type quarks and down-type quarks separately are common with the Peccei-Quinn(PQ) symmetry \cite{PQ77,KimRMP10} and with SUSY. The GUT scale singlets $X_{\pm 1}^{d,u}$ are the FN fields. We do not introduce $X_{\pm m}^{d,u}$ for $2\le m\le 5$ in the hope that the expansion parameter in each quark mass texture is just one parameter $|X_{\pm 1}^{d,u}|$. In Table \ref{tab:Discrete}, we present $\Z_{12}$ quantum numbers of these fields.  In the table, in addition we also presented the fields $X_6^{d,u}$ and $X_0^{d,u}$ which are needed in the superpotential to determine $\delta=\frac{\pi}{2}$.

As commented before, with SUSY the needed Yukawa couplings can be written with the set-and-forget principle. But, we introduce $\Z_4$ and $\Z_3$ without invoking the set-and-forget principle. In Table \ref{tab:Discrete}, we present  $\Z_{12}, \Z_4$ and $\Z_3$ quantum numbers of left-handed quark doublets, right-handed quark singlets, Higgs doublets, the FN fields $X_{\pm 1}^{d,u}$, and the SM singlet fields $X_{6,0}^{d,u}$. The SM singlet fields $X_{6,0}^{d,u}$ are needed for generating the needed VEVs. $X_0^{d,u}$ is expected to be of order 1. $X_{\pm 1}^{d,u}$ is of order $\lambda$, and $X_{6}^{d,u}$ is of order $\lambda^6$.  The $\Z_4$ and $\Z_3$ symmetries are needed to keep the leading terms of Eq. (\ref{eq:TextureRL}). These guarantee the vanishing entries of the quantum number elements of $\tilde M^{(d)}$ in Eq. (\ref{eq:TextureRL}). With $\Z_4$ and $\Z_3$, the up-type $X^u_{\pm 1}$ and the down-type $X^d_{\pm 1}$ couple to $Q_{\rm em}=\frac23$ quarks and $Q_{\rm em}=-\frac13$ quarks separately.
One can obtain the transformation properties of the fields under $\Z_4$ and $\Z_3$ from Table \ref{tab:Discrete}.
\begin{widetext}
%%%%%%%%%%%%%%%%%%%%%%%%%%%%%%%%%%%%%%%%%%%%%%%%%%%%%%%%%%%%%%%%%%%%%%%
\begin{table}
\begin{center}
\begin{tabular}{c|ccc|ccc|ccc|cc|cccc|cccc}
 &$\overline{q}_{1L}$& $\overline{q}_{2L}$& $\overline{q}_{3L}$& $d_R$& $s_R$& $b_R$ & $u_R$ & $c_R$ & $t_R$ & $H_d$& $H_u$ & $X_1^{d}$ & $X_{-1}^{d}$ &$X_1^{u}$ & $X_{-1}^{u}$ & $X_6^{d}$&$X_0^{d}$ &$X_{6}^u$ &$X_0^{u}$  \\[0.3em] \hline
 $\Z_{12}$ &+1 &0 & $-2$   & $-5$&  0 & +2 & +5& +4&+2 & 0 & 0 & +1  & $-1$ &+1& $-1$
 &6 &0 &6 & 0 \\
\hline
$\Z_4$ & 2& 0& 0& 0& 0& 0& 2&0&0& $2$&0& $1$& $3$& $2$& $2$& 2& 2& 0& 0\\
$\Z_3$ & 0&1&0&2&2& 1&   0&1&0&  1&0& $0$&$0$&$2$&$1$&  0&2&0 &0
%\hline
\end{tabular}
\caption{The $\Z_{12}$ charges of the fields. There can be additional $\Z_4$ and $\Z_3$ symmetries.  The indices $d$ and $u$ denote the coupling to the right-handed $d$ and $u$ quarks, respectively.
}
\label{tab:Discrete}
\end{center}
\end{table}
%%%%%%%%%%%%%%%%%%%%%%%%%%%%%%%%%%%%%%%%%%%%%%%%%%%%%%%%%%%%%%%%%%%%

Then, the up and down type quark mass matrices are given, only for the leading term in each element, as
\dis{
 \tilde M^{(u)}&=\left(\begin{array}{c|ccc}  & u_R(+5)  & c_R(+4)  & t_R(+2)\\[0.3em] \hline
 \overline{q}_1(+1)    & cX_{-1}^{u\,6}    & -c X_{-1}^{u\,5} & \kappa_t X_{-1}^{u\,3} \\[0.2em]
  \overline{q}_2(0)    & -cX_{-1}^{u\,5}  &  c X_{-1}^{u\,4}  & -\kappa_t X_{-1}^{u\,2} \\[0.2em]
      \overline{q}_3(-2)    & \kappa_t X_{-1}^{u\,3}   &  -\kappa_t X_{-1}^{u\,2} & 1
\end{array}\right) v_u\, ,\\[0.4em]
  \tilde M^{(d)}&=\left(\begin{array}{c|ccc}  & d_R(-5)  & s_R(0)  & b_R(+2)\\[0.3em] \hline
 \overline{q}_1(+1)    & dX_{+1}^{d\,4}   &  0 & 0\\[0.2em]
  \overline{q}_2(0)    & 0   &  s  X_{+1}^d X_{-1}^d  & \kappa_b X_{-1}^{d\,2} \\[0.2em]
      \overline{q}_3(-2)    & 0   & \kappa_b X_{+1}^{d\,2} & 1
\end{array}\right) v_d\,.             \label{eq:TexMuMd}
}
\end{widetext}
where the appropriate powers of $M_P^{-1}$ are multiplied to make the elements of the mass dimension. Note the parameter $u$ (the coefficient of the mass eigenvalue of the up quark in Eq. (\ref{eq:MuMd})) does not appear in the leading terms in Eq. (\ref{eq:TexMuMd}).

The specific forms $\tilde M^{(u)}$ and $\tilde M^{(d)}$ of (\ref{eq:TextureRL}) are obtained requiring Arg.Det.$V_{\rm CKM}=0$, which however is not a physically required condition. Changing this condition allows two unobservable quark phases. If Det.$\tilde M^{(u)}$ has a phase, then the Det.$\tilde M^{(u)}$ phase can be removed by redefining $u_R,c_R,$ and $t_R$ each absorbing the third of the Det.$\tilde M^{(u)}$ phase. Similarly, if  Det.$\tilde M^{(d)}$ has a phase, then that phase also is removed by redefining $d_R,s_R,$ and $b_R$ each absorbing the third of the Det.$\tilde M^{(d)}$ phase. Therefore, even though the form (\ref{eq:TextureRL}) is derived from the useful CKM matrix \cite{KimSeo11}, we must allow two overall phases, one in $\tilde M^{(u)}$ and the other in $\tilde M^{(d)}$. In this paper, however, this possibility is not needed.

Note that the (33) elements of $\tilde M^{(u)}$ and $\tilde M^{(d)}$ in Eq. (\ref{eq:TexMuMd}) are set to 1 since their $\Z_{12}$ quantum numbers are zero. The (33) element of $\tilde M^{(u)}$ satisfies the $\Z_4$ and $\Z_3$ discrete symmetries also. However, the (33) element of $\tilde M^{(d)}$  does not satisfy the $\Z_4$ and $\Z_3$ discrete symmetries. Here, we use the discrete gauge symmetry idea that the Planck scale physics destroys the discrete symmetry if it is not a subgroup of a gauge symmetry \cite{Krauss89}. Except the (33) element of  $\tilde M^{(d)}$, the matrices in Eq. (\ref{eq:TexMuMd}) describe the terms respecting the SM gauge and $\Z_{12}\times \Z_4\times \Z_3$ discrete symmetries. The Planck scale physics would lead to a democratic form of a mass matrix, even though it does not respect the discrete symmetries. To keep all other terms respect the symmetries of Table \ref{tab:Discrete}, we assign 1 at the (33) position of  $\tilde M^{(d)}$. Under this philosophy, we assume that the matrices are proportional to a democratic form \cite{FritzschDemo90},
\dis{
\left(\begin{array}{ccc}
\frac13 & \frac13  & \frac13  \\ \frac13  & \frac13  & \frac13  \\ \frac13 & \frac13 & \frac13
 \end{array}  \right),\nonumber
}
which gives one massive quark and two massless quarks with the quark mass matrix in the new basis becomes
\dis{
\left(\begin{array}{ccc}
0 & 0 & 0 \\ 0 & 0 & 0 \\ 0 & 0 & 1
 \end{array}  \right).\label{eq:DemoMass}
}
Note in particular that the determinant having two zero eigenvalues must be satisfied.
An important lesson from this is that we must have a correct traces and determinants of the matrices. For the up-type quarks the trace must be $1+{\cal O}(\lambda^4)+{\cal O}(\lambda^6)$, and for the down-type quarks the trace must be $1+{\cal O}(\lambda^2)+{\cal O}(\lambda^4)$. These trace conditions are read from mass textures, but not from the mass eigenvalues. For example, for the up quark sector, even though the trace of mass eigenvalues is $m_t+m_c+m_t(u\lambda^7)$, we keep the trace only up to $\lambda^6$ whose order is a correction to $m_t$ and $m_c$. Since there are two zero eigenvalues for the democratic form (\ref{eq:DemoMass}), the following $2\times 2$ submatrix (in the lower right corner) condition is also satisfied,
\dis{
\left(\begin{array}{cc}
  0 & 0 \\ 0 &  1
 \end{array}  \right).\label{eq:twobytwo}
}
With this understanding, below $M_P$ we take the Planck scale contribution to mass matrices as Eq. (\ref{eq:DemoMass}). If we add more ${\cal O}(1)$ terms to Eq. (\ref{eq:DemoMass}) due to gravity, we cannot satisfy the $3\times 3$ matrix condition Eq. (\ref{eq:DemoMass}) and the $2\times 2$ submatrix condition  Eq. (\ref{eq:twobytwo}). This is the logic we insert 1 in the (33) elements of Eq. (\ref{eq:TexMuMd}) even if those (33) entries violate the symmetries of Table \ref{tab:Discrete}. But all the other entries are required to respect the symmetries. Namely, the discrete symmetry violation by gravity is moved to the (33) entries only, in fact to one position in the present example: the (33) entry of  $\tilde M^{(d)}$.

%%%%%%%%%%%%%%%%%%%%%%%%%%%%%%%%%%%%%%%%%%%%%%%%
\section{Maximal CP violation}\label{sec:MaxCP}
For a calculable CP phase, we start with real couplings in the Lagrangian, which is usually adopted in calculable $\bar\theta$ models. With SUSY all the parameters of the superpotential $W$ are real, and the initial QCD vacuum angle $\theta_{QCD}$ is zero \cite{KimRMP10}.

Terms including $X_0^{(d,u)}$ are
\dis{
W^{(0)}=&~ X_0^{(u)}(\alpha X_{+1}^{(d)}X_{-1}^{(d)} +\tilde\alpha X_{+1}^{(u)}X_{-1}^{(u)})
+ \tilde m_0^2 X_0^{(u)}  \\
&+\frac{\alpha'}3 X_0^{(u)3}-\tilde M_0 X_0^{(u)2}\\
   &+\frac{h_{00}^3}{6M_P^3} X_0^{(d)6} +\frac{h_{06}^2}{3M_P^2} X_0^{(d)3}X_6^{(d)}X_6^{(u)}.
   \label{eq:WX0}
}
In Eq. (\ref{eq:WX0}), we will take the limit $\tilde m_0^2\to 0$.
\dis{
\frac{dW^{(0)}}{dX_{0}^{(u)}}=&\alpha X_{+1}^{(d)}X_{-1}^{(d)} +\tilde\alpha X_{+1}^{(u)}X_{-1}^{(u)} -2\tilde M_0 X_{0}^{(u)}\\
&+\alpha'  X_0^{(u)2}+\tilde m_0^2 =0\\
\frac{dW^{(0)}}{dX_{0}^{(d)}}=& \frac{h_{00}^3}{M_P^3} X_{0}^{(d)5} +\frac{h_{06}^2}{M_P^2} X_0^{(d)2}X_6^{(d)}X_6^{(u)} =0\label{eq:SUSYcond}
}
In the limit $\tilde m_0^2\to 0$, $X_{0}^{(u)}$ is determined as
\dis{
X_{0}^{(u)}\simeq \frac{\tilde M_0}{\alpha'}\pm \sqrt{\left(\frac{\tilde M_0}{\alpha'}
\right)^2-  \frac{\alpha X_{+1}^{(d)}X_{-1}^{(d)} +\tilde\alpha X_{+1}^{(u)}X_{-1}^{(u)}}{\alpha'}  }.\nonumber
}
%In the limit of $\frac{\tilde M_0}{\alpha'}\sim M_P$, the smaller eigenvalue of $X_{0}^{(u)}$ is
%\dis{
% X_{0}^{(u)}\simeq \frac{\alpha X_{+1}^{(d)}X_{-1}^{(d)} +\tilde\alpha X_{+1}^{(u)}X_{-1}^{(u)}}{2 \tilde M_0 }\,.
% }
%If $\alpha=-\tilde\alpha,\, \alpha'=0$ and $\tilde M_0=0$ are assumed, we obtain from Eq. (\ref{eq:SUSYcond}) the same expansion parameter $\lambda$ for $\tilde M^{(u)}$ and $\tilde M^{(d)}$. 
The second equation of (\ref{eq:SUSYcond}) leads to
\dis{
X_0^{(d)}= &\left(-\frac{h_{06}^2 X_6^{(d)}X_6^{(u)} M_P}{h_{00}^3 }\right)^{1/3}\\
&\approx \lambda^4\left(-\frac{h_{06}^{2/3}}{h_{00} } \right)  M_P \gtrsim \lambda M_P,
}
where we assumed $X_6^{(d,u)}\simeq \lambda^6 M_P$, and we allow the possibility of a smaller $m_s$ compared to $m_\mu$ \cite{GeoJarlskog79}.
So, we need a fine-tuning of couplings  $|{h_{00}}/{h_{06}^{2/3}}|  \lesssim\lambda^3\sim 10^{-2}$.

The needed weak CP violation occurs through developing complex vacuum expectation values \cite{LeeTD73}. The symmetries of Table \ref{tab:Discrete} allow the following down-type singlets terms in $W$,\footnote{To simplify the notations, for the $W^{(d)}$ discussion we suppress the superscript $d$ of the down-type singlets, and for the $W^{(u)}$ discussion we suppress the superscript $u$ of the up-type singlets.}
\dis{
W^{(d)}= &\frac12 \mu_6 X_{6}X_{6}  +\mu_1 X_{+1}X_{-1}\\
&+\frac{c_{11}}{2M_P}X_{+1}^2X_{-1}^2 +\frac{c_{66}}{4M_P}X_{6}^4 +W' \\
W'=&\frac{f}{M_P^4}X_{6}X_{-1}^6+\frac{g}{M_P^4}X_{6}X_{+1}^6
\label{eq:Veffdown}
}
where $\mu_1$ and $\mu_2$ are real and $g=\pm f$  and $c_{ij}$ are real. There are more dimension 4 terms including $X_{6}^2$ and $X_{+1}X_{-1}$ which are neglected for simplicity, since they do not change our introduction of the phase $\frac{\pi}{2}$. We introduced the mixing term $W'$ which relates the phases of $X_6$ and $X_{+1}$. We can multiply $X_0^{(u)}$  to any of these terms which however is assumed to be smaller.
The SUSY points for $X_{\pm 1}$ and $X_{+6}$ fields are given by,
\begin{widetext}
\dis{
&\frac{dW^{(d)}}{dX_{-1}}X_{-1}=\mu_1 X_{+1}X_{-1}+6\frac{f}{M_P^4}X_{+6} X_{-1}^6+\frac{c_{11}}{M_P}X_{+1}^2 X_{-1}^2=0\\
&\frac{dW^{(d)}}{dX_{+1}}X_{+1}=\mu_1 X_{+1}X_{-1}+6\frac{g}{M_P^4}X_{+6} X_{+1}^6+\frac{c_{11}}{M_P}X_{+1}^2 X_{-1}^2=0\\
&\frac{dW^{(d)}}{dX_{+6}}=\mu_6 X_{+6}+\frac{f}{M_P^4} X_{-1}^6+\frac{g}{M_P^4}  X_{+1}^6 +\frac{c_{66}}{M_P}X_{+6}^3=0\label{eq:VEVsX1}
}
\end{widetext}
For $g=f$, we do not obtain a phase for $X_{\pm 1}^2$. For $g=-f$, we obtain from the first two equations,
\dis{
X_{+1}^6+X_{-1}^6 &=0,~~ X_{\pm 1}=\left(\frac{1}{\sqrt2} \pm i\frac{1}{\sqrt2}\right)\left| X_{\pm1} \right|\,,\\
&X_{\pm 1}^2=\pm i |X_{\pm 1}|^2\label{eq:pio4}
}
and
\dis{
6\frac{f}{M_P^4}I_6|X_{+1}|^4-\frac{c_{11}}{M_P}|X_{+1}|^2-\mu_1=0 \label{eq:AbsX1}
}
where $X_6=R_6 +iI_6$, \ie  $X_{+6}$ is pure imaginary  $X_6=iI_6$.
From Eq. (\ref{eq:AbsX1}), we determine the smaller $|X_{+1}|^2$ as
\dis{
\frac{|X_{+1}|^2}{M_P^2}&=\frac{c_{11}M_P}{12 fI_6}-\sqrt{\left(\frac{c_{11}M_P}{12 fI_6} \right)^2 +\frac{\mu_1}{6 fI_6}}\\
&\simeq -\frac{\mu_1}{c_{11}M_P}\simeq -\lambda^2\,,\label{eq:lambda2}
}
where $\mu_1$ and $c_{11}$ are tuned to satisfy Eq. (\ref{eq:lambda2}).
From the third equation of (\ref{eq:VEVsX1}), we have
\dis{
\left(\frac{\mu_6}{c_{66}M_P}\right)\frac{I_6}{M_P}+ \frac{2f}{c_{66}}\lambda^6-\left(\frac{I_6}{M_P} \right)^3=0
}
which has a solution $I_6\simeq  M_P\sqrt{ \mu_6/c_{66}M_P}$ in the limit $\lambda^6\to 0$. However, the solution $I_6\sim \lambda^6$ is the one we need so that a small expansion parameter is $X_{\pm 1}$ and any expansion parameter involving $I_6$ must be of order $\lambda^6$ and smaller. For $I_6\to 0$, we have,
\dis{
\frac{I_6}{M_P}\simeq -\left(\frac{2f M_P}{\mu_6} \right)\lambda^6.\label{eq:I6mag}
}

This solution leads to the maximal weak CP violation since the phase $\delta$ appearing in Eq. (\ref{eq:TexMuMd}) is the phase of $X_{+1}^2$ which is $\frac\pi{2}$, viz. Eq. (\ref{eq:pio4}). But the maximal CP violation is completed with obtaining an appropriate up-type quark mass texture of Eq. (\ref{eq:TexMuMd}).

For the up-type Higgs singlets, we need the real VEVs except the phase leading to an overall phase of \Vckm. The superpotential for the $Q_{\rm em}=+\frac23$ quarks is\footnote{Note that for the $W^{(u)}$ we suppress the superscript $u$ of the up-type singlets.}
\dis{
W^{(u)}= &\frac12\tilde \mu_6 X_{6}X_{6}  +\tilde \mu_1 X_{+1}X_{-1}\\
&+\frac{\tilde c_{11}}{2M_P}X_{+1}^2X_{-1}^2 +\frac{\tilde c_{66}}{4M_P}X_{6}^4 +W' \\
\tilde W'=&\frac{\tilde f}{M_P^4}X_{6}X_{-1}^6+\frac{\tilde g}{M_P^4}X_{6}X_{+1}^6
\label{eq:Veffup}
}
where $\tilde \mu_1$ and $\tilde \mu_2$ are real and $\tilde g=\pm \tilde f$  and $\tilde c_{ij}$ are real. We introduced the mixing term $\tilde W'$ which relates the phases of $X_6$ and $X_{+1}$.

The SUSY points for $X_{\pm 1}$ and $X_{6}$ fields are given by,\\

\begin{widetext}
\dis{
&\tilde \mu_1 X_{+1}X_{-1}+6\frac{\tilde f}{M_P^4}X_{6} X_{-1}^6+\frac{\tilde c_{11}}{M_P}X_{+1}^2 X_{-1}^2=0\\
&\tilde \mu_1 X_{+1}X_{-1}+6\frac{\tilde g}{M_P^4}X_{6} X_{+1}^6+\frac{\tilde c_{11}}{M_P}X_{+1}^2 X_{-1}^2=0\\
&\tilde \mu_6 X_{6}+\frac{\tilde f}{M_P^4} X_{-1}^6+\frac{\tilde g}{M_P^4}  X_{+1}^6 +\frac{\tilde c_{66}}{M_P}X_{6}^3=0\label{eq:VEVsX}
}
\end{widetext}
Not to have a phase, we choose $\tilde g=\tilde f$. Then, we obtain the real solution, $X_6=R_6$ and $X_{\pm 1}=R_1$, and obtain an equation,
\dis{
\frac{R_1^4}{M_P^4} + \frac{\tilde{c}_{11}}{6\tilde fR_6 M_P}R_1^2+\frac{\tilde\mu_1 }{6\tilde fR_6} =0\,,\label{eq:upR1}
}
which leads to a smaller solution of $R_1^2$ as
\dis{
\frac{R_1^2}{M_P^2}&=- \frac{\tilde c_{11}M_P}{12\tilde fR_6}+ \sqrt{\left(\frac{\tilde{c}_{11}M_P}{12 \tilde fR_6} \right)^2 -\frac{\tilde\mu_1}{6 \tilde fR_6}}\\
&\simeq -\frac{\tilde \mu_1}{\tilde c_{11}M_P}\simeq -\tilde\lambda^2\,,\label{eq:smallR1}
}
where $\tilde \mu_1$ and $\tilde c_{11}$, having the opposite signs, are tuned to satisfy Eq. (\ref{eq:lambda2}). As in the down-type case, $R_6$ is determined from a cubic equation. To have a universal $\lambda$, \ie $\lambda$ of Eq. (\ref{eq:lambda2}) and $\tilde\lambda$ of Eq. (\ref{eq:smallR1}) the same, we need $\tilde\mu_1/\tilde{c}_{11}= \mu_1/{c}_{11}$.

The $\Z_{12}$ symmetry has $\Z_4$ and $\Z_3$ as its subgroups. So, the down-type symmetry $\Z_4$ can allow the phase $\frac{\pi}{2}$.
The chief merit of the $\Z_{12}$ symmetry is to introduce the FN type powers of $\lambda$.

The above determination of the CKM matrix with the maximal CP phase needs to be completed in a more complete theory. We present two speculations related to our determination of the CKM matrix:
\begin{itemize}
\item
The discrete symmetries of Table \ref{tab:Discrete} allow the following superpotential term,
\dis{
 X_0^{(d)}  H_uH_d
\sim  M_P H_uH_d  \label{eq:mu}
}\\
which gives a too large $\mu$-term. Therefore, we need to introduce a PQ symmetry \cite{PQ77} to suppress the $\mu$-term of (\ref{eq:mu}).
If it were not for this unacceptably large $\mu$ of Eq. (\ref{eq:mu}), this model is a good example of the calculable $\bar{\theta}$ \cite{KimRMP10} because the higher order corrections do not destroy the hermiticity nature of $\tilde M^{(d)}$.

\item We speculate that the parameters $c,d,s, \kappa_t$, and $\kappa_b$ of Eq. (\ref{eq:TexMuMd}) are determined in an ultraviolet completed theory, probably by geometrical factors.

\item There are a few unsatisfactory features in the present example. For example, firstly the equivalence of $\lambda$ of  Eq. (\ref{eq:lambda2})  and $\tilde\lambda$  of Eq. (\ref{eq:smallR1}) requires a fine tuning. Second, the initial flavor democratic mass matrix is redefined such that the discrete symmetry violating entry, the (33) element of $\tilde M^{(d)}$, absorbs the nonzero mass eigenvalue. We hope that they can be explained in a better model.\\

\end{itemize}

%%%%%%%%%%%%%%%%%%%%%%%%%%%%
\section{Conclusion}\label{sec:Conclusion}
Due to the accurate determination of the CKM parameters, it is possible to obtain the quark mass matrices fairly reliably. For the same left-hand unitary matrices $L^{(u),(d)}$ and the right-hand unitary matrices $R^{(u),(d)}$, \ie $L^{(u),(d)}=R^{(u),(d)}$, the weak-basis quark-mass matrices take simple forms.  In contrast to continuous symmetries, with discrete symmetries the CP phase of $2\pi$ divided by an integer can be obtained from a discrete symmetry since the degenerate vacua of discrete symmetries are countable. Introducing a $\Z_{12}$ symmetry, we obtain the maximal CP phase $\frac{\pi}{2}$.  We introduced a $\Z_{12}$ discrete symmetry with the FN scalars since the  the lightest element($\bar uu$ element) of the quark mass matrices has a power $\lambda^6$. We considered the $\Z_{12}$ symmetry at field theory level. With SUSY, we can follow the set-and-forget principle for the needed and forbidden terms in the superpotential. Barring the set-and-forget principle, however, we need additional $\Z_4\times \Z_3$ symmetries to obtain the desired superpotential.   In this paper, we have shown the maximality of CP violation with SUSY but the maximality might be obtained also without SUSY if we introduced an appropriate discrete symmetry. Finally, we note that the $\Z_{12}$ symmetry may have a root in string theory such as in a $\Z_{12}$ orbifold compactification \cite{KimKaye07}.

%%%%%%%%%%%%%%%%%%%%%%%%%%%%%%%%%%%%%%%%%%%%%%%%%%%%%%%%%%%%%%%%%%%%%%%%%%%%
\acknowledgments{I thank B. Kyae and M. Seo for helpful discussions. This work is supported in part by the National Research Foundation  (NRF) grant funded by the Korean Government (MEST) (No. 2005-0093841).}

\vskip 0.5cm
%%%%%%%%%%%%%%%%%%%%%%%%%%%%%%%%%%%%%%%%%%%%%%%%%%%%%%%%%%%%%%%%%%%%%%%%%%

\end{document}